\documentclass[a4paper,11pt]{article}

\pdfoutput=1 % if your are submitting a pdflatex (i.e. if you have
\usepackage{jcappub} % for details on the use of the package, please
\usepackage[T1]{fontenc} % if needed

\title{Direction-of-arrival estimation of a gravitational wave by correlations between quadrupole moments 
of pulsar timing data}

\author[a]{Taichi Ueyama}
\author[a]{Hodaka Tamura}
\author[a]{Hideki Asada} 

\affiliation[a]
{Graduate School of Science and Technology, 
Hirosaki University,
Hirosaki 036-8561, Japan}

\emailAdd{h24ms103@hirosaki-u.ac.jp}
\emailAdd{h25ms122@hirosaki-u.ac.jp}
\emailAdd{asada@hirosaki-u.ac.jp}

\abstract{
Can we estimate the direction of arrival (DOA) of a gravitational wave (GW) signal 
from pulsar timing array observations? 
The present paper addresses the inverse problem, 
for which 
we consider quadrupole moments of pulsar timing data 
due to GWs from a dominant isolated source such as a binary of supermassive black holes 
over an isotropic stochastic background. 
Correlations between the quadrupole moments are discussed 
as a possible alternative to the matched filtering technique, 
where the correlations between pulsar pairs over 
the full sky are taken into account.  
The correlations turn out to be in the form of  
a three-dimensional traceless matrix with 
rank 3 that can be closely related with a projection tensor for the GW. 
Thereby, we demonstrate that the rank-3 matrix allows to estimate 
the DOA of the GW 
without assuming a template bank nor a priori Gaussian noise. 
A mean over the celestial sphere is adopted instead of the pulsar averaging. 
A weighted average is used for a non-uniform distribution of pulsars on the sky, 
for which the Voronoi tessellation is employed to numerically calculate an area-weighted mean. 
In expectation of the forthcoming Square Kilometre Array, 
DOA estimation errors are also examined. 
}

\arxivnumber{2602.21607}

\begin{document}
\maketitle
\flushbottom

\section{Introduction}
The idea of a search for gravitational waves (GWs) 
by using radio pulse timings 
can be dated back to Reference 
\cite{Estabrook, Sazhin, Detweiler}. 
In their pioneering work, 
Hellings and Downs (HD) 
found that the sky correlation pattern of pulse timings 
among pulsar pairs depends upon the angle $\gamma$ 
between the lines of sight to the pulsars viewed from Earth, 
and the correlation pattern can be a strong evidence of GWs 
\cite{Hellings}, 
because it reflects the quadrupole nature of GWs 
\cite{CreightonBook, MaggioreBook, Anholm2009, Jenet}. 
See also Reference 
\cite{Romano,Romano2024} 
for a review on detection methods of stochastic GW backgrounds. 

Eventually, 
several teams of pulsar timing arrays (PTAs) have 
reported a strong evidence of nano-hertz GWs 
\cite{Agazie2023, Antoniadis2023, Reardon2023, Xu2023}. 
According to these papers, 
a superposition of supermassive black hole binaries (SMBHBs)
are among possible GW sources for the nano-hertz GWs, 
though improved methods, other possible explanations, and new physics searches 
have been widely argued 
e.g. \cite{Chen2020, Ellis, Smarra, Gouttenoire, Franciolini2023, DeRocco, Franciolini2024, Figueroa, Athron, Shih, Kumar, Bian, Chen2024, Liu2023, Liu2024}. 
A possible variance around the HD correlation as the mean has been discussed 
\cite{Allen2023, Allen2024}.

If a single SMBHB dominates in a certain frequency bin 
for PTA observations, 
can we estimate the direction of arrival (DOA) for the GW? 
Currently,  this issue is theoretical. 
See e.g. \cite{Finn2010, Burt2011, Becsy} 
for methods for a single GW source detection by using PTAs.  

Yet, it is expected that the near future Square Kilometre Array (SKA) 
will drastically increase 
the number of pulsars that can be used to significantly improve the sensitivity of PTAs 
\cite{SKA}. 
Sasaki et al. have recently proposed the use of a hemisphere in stead of the full sky 
in calculating the pulsar timing correlations 
\cite{Sasaki2024}. 
Introducing a hemisphere into PTAs violates the spatial isotropy, 
such that hemisphere-averaged correlations can have a dependence on the DOA. 
However, there is a drawback in the hemisphere method, because the number of pulsars 
is reduced to nearly a half, resulting in a lower detection sensitivity. 

Can there exist a method of estimating the DOA of GWs 
by using the whole of the observed pulsars? 
This issue is not new and it has been extensively studied 
by using the matched filtering technique, 
which is derived from assumptions of the Gaussian noise. 
Therefore, the matched filtering is expected to be optimal in the presence of Gaussian noise 
\cite{Babak, Ellis2012}. 
However, it is computationally expensive because 
it needs a huge template bank 
for PTA data sets including messy noises 
\cite{CreightonBook, MaggioreBook, Agazie2023b, Chalumeau}. 
See e.g. \cite{Taylor, Kato, Jow, Schult, Tsai} for numerical approaches 
based on the matched filtering.

There has been a work on localizing binaries on the sky, 
especially in the papers \cite{Lee, Rosado, Petrov, Taylor2026, Kopeikin}. 
The PTA response function to GWs from an individual SMBHB 
is examined to calculate the accuracy of determining the GW properties 
especially by taking account of the pulsar term 
\cite{Lee}, 
in which the accuracy of pulsar distance measurement from the strong GW signal 
is discussed. 
Expected properties of the first GW signal with PTAs are argued 
\cite{Rosado},  
where the pulsar terms are included. 
Toward identifying the host galaxy of a SMBHB by using PTAs, 
GW effects are numerically modeled by including the pulsar term 
\cite{Petrov}, 
which suggests that the expected area of the single GW source 
is around a hundred square degrees for a high signal-to-noise ratio case. 
A comprehensive study of the antenna pattern function including a phase difference 
between the Earth term and  the pulsar one is done 
\cite{Taylor2026}, 
in which 
the angular localization is examined. 
In particular, a possible influence of pulsar distance uncertainty 
is argued. 
The Earth-term-dominated regime and the pulsar-term-dominated regime 
have been extensively studied. 
The transition regime between the two dominated ones has been 
recently investigated by analyzing a mathematical structure of 
the detector response function 
through a tensor spherical harmonic decomposition of the gravitational-wave field 
\cite{Kopeikin}.

For addressing the above issue by using an alternative method without 
assuming a priori Gaussian noise, 
the main purpose of the present paper is to 
discuss correlations between 
quadrupole moments of pulsar timing data 
over the full sky as well as 
over the whole observation time. 
The quadrupole-moment correlation method 
does not need a huge template bank, 
because it does not make fittings with templates.   
We shall discuss also how to estimate the DOA 
from the quadrupole correlations. 
Therefore, the quadrupole method could serve as 
an independent consistency check for the position of a single SMBHB 
after its initial detection is done probably by the matched filtering technique.

This paper is organized as follows. 
Section 2 discusses quadrupole moments of the pulsar timing data, 
where the correlations between the quadrupole moments are 
taken over the observed pulsar pairs as well as over the whole observation time. 
Section 3 studies a relationship among the correlations, a GW projection tensor, and a DOA estimation of GWs. 
Section 4 examines estimation errors by random noises. 
Section 5 is devoted to Summary. 
Appendix A provides a few steps for deriving Eq. (\ref{Integral-F-q}), 
which plays a key role in this paper. 
Appendix B shows some relations between the GW polarization tensors, 
which are used in Section 3. 
Throughout this paper, 
the unit of $c=1$ is used and 
the center of the coordinates is chosen as the solar barycenter, 
safely approximated as the Earth.

%%%%%
\section{Correlations  of quadrupole tensors}
\subsection{Setup and notations}
We suppose that $N_p$ of pulsars are 
observed by a pulsar timing array. 
A pair of pulsars are labeled by $a$ and $b$ 
for $a, b \in (1, 2, \cdots, N_p)$, 
where the directions of the $a$-th and $b$-th pulsars 
are denoted by unit vectors 
$\hat\Omega_a$ and $\hat\Omega_b$, 
respectively.

The observed signals for the $a$-th and $b$-th pulsars are 
\cite{CreightonBook, MaggioreBook, Anholm2009, Jenet}
\begin{align}
s_a(t) 
&= 
z_a(t) + n_a(t) , 
\label{sa}
\\
s_b(t) 
&= 
z_b(t) + n_b(t) , 
\label{sb}
\end{align}
where 
$z_a(t)$ and $z_b(t)$ 
mean redshifts of a radio pulse due to GWs, and  
$n_a(t)$ and $n_b(t)$ are noises. 
We focus only on the Earth terms. 
The present paper neglects pulsar terms, 
because they behave as random phases 
for a large number of pulsars whose precise distances are not known 
and they are thus negligible in the average 
when the characteristic GW wavelength is much shorter than pulsar distances 
e.g \cite{Hellings, CreightonBook, MaggioreBook, Cornish, Romano2024}. 

The current work considers a situation that 
the Earth term is dominant when a pulsar distance $D_p$ is much larger than 
the wavelength $\lambda_{GW}$ of GWs under study, 
such that pulsar terms can be ignored.  
When $D_p \sim \lambda_{GW}$, 
on the other hand, 
an interference between the Earth term and the pulsar term 
becomes significant and a PTA response is largely affected, 
as illustrated by e.g.  Figure 2 of Reference \cite{Lee}. 
The detection probability and a parameter space have been numerically discussed 
\cite{Rosado,Petrov}. 

There is an exceptional case where the pulsar term plays a crucial role even in 
the Earth-term-dominated regime. 
If a few pulsars with a distance precision of 1 pc are observed, 
the precision of the source location will be improved by 
the use of the pulsar term 
\cite{Kato}. 
 See also \cite{Taylor2026} for a comprehensive study of these issues.

Let us imagine that a signal in a certain frequency bin of PTA observations 
is dominated by a distant SMBHB in circular motion 
and in the direction $\hat\Omega_{GW}$ from the Earth. 
In the theory of general relativity, 
the redshifts are written as 
\cite{CreightonBook, MaggioreBook, Anholm2009, Jenet}
\begin{align}
z_a(t) 
&= 
h^{+} F^{+}_{a} 
\cos(\omega t) 
+  
h^{\times} F^{\times}_{a} 
\sin(\omega t) , 
\label{za}
\\
z_b(t) 
&= 
h^{+} F^{+}_{b} \cos(\omega t) 
+  
h^{\times} F^{\times}_{b}\sin(\omega t) ,  
\label{zb}
\end{align}
where $h^{+}$ and $h^{\times}$ denote constants for amplitudes of 
the plus and cross modes, respectively, 
and 
the initial phase of GWs can be zero by choosing an initial time. 
Here, the GW antenna patterns in the plane-wave approximation 
are defined as 
\cite{CreightonBook, MaggioreBook, Anholm2009, Jenet}
\begin{align}
F_a^A
\equiv \frac{1}{2} \frac{\hat\Omega_{a}^i \hat\Omega_a^j}
{1+\hat\Omega_{GW} \cdot \hat\Omega_a} e_{i j}^A(\hat\Omega_{GW}) ,
\label{F}
\end{align} 
where 
$e_{i j}^A(\hat\Omega_{GW})$ denotes the polarization tensor along the direction $\hat\Omega_{GW}$ 
for $A = +, \times$. 
See References \cite{McGrath2021,McGrath2022,Guo,Kubo,D'Orazio} 
for possible corrections by a nearby SMBHB and their cosmological implication. 
See also \cite{Hu} 
for discussions on different frequencies emerging from the same GW source 
in a search of new physics beyond general relativity. 
Beat effects can occur in the case that there are two SMBHBs that produce 
two GWs with very similar frequencies and comparable amplitudes at the Earth. 
A detection method for the beat effects has been discussed 
\cite{Yamamoto2025}. 
The present paper considers neither a gravity test nor a beat effect.

\subsection{Correlations of quadrupole moments of pulsar timing data}
A quadrupole and traceless part of the redshifts over the sky can be defined as 
\begin{align}
I^{ij}(t) 
\equiv 
\frac{1}{4\pi} \oint d\Omega_a z_a(t) q_a^{ij}  , 
\label{I}
\end{align}
where $\oint d\Omega_a$ is an integral over the full sky, 
and 
\begin{align}
q_a^{ij} 
\equiv 
\hat\Omega_{a}^i \hat\Omega_a^j - \frac13 \delta^{ij} . 
\label{omega}
\end{align}

The present paper uses the averaging 
over the celestial sphere as Eq. (\ref{I}), 
which differs from the pulsar averaging \cite{Cornish}.  
We refer to it as the celestial sphere averaging. 

The pulsar averaging is a mean over a pulsar distribution in the sky. 
When the actual pulsar distribution is anisotropic, 
a quadrupole (and higher) moment due to the anisotropic distribution 
can appear but it is unrelated to the quadrupole nature of GWs. 
Hence, the pulsar averaging is not suitable for the purpose of estimating the DOA.  

On the other hand, 
the celestial sphere averaging is a mean over the celestial sphere, 
defined by Eq. (\ref{I}), 
which is unrelated to an anisotropy in the pulsar distribution. 
It coincides with the pulsar averaging, if the pulsar distribution is isotropic. 
However, the two averaging methods are usually different from each other.

A naive intuition may mislead to a candidate for 
correlations of quadrupole moments in pulsar timing data as 
\begin{align}
\frac{1}{T_{obs}}
\int_0^{T_{obs}} dt 
\left[
\left(
\frac{1}{4\pi}
\oint d\Omega_a 
%\left(
\Bigl(
s_a(t) q_a^{ij} 
\Bigr)
%\right)
\right)
\left(
\frac{1}{4\pi}
 \oint d\Omega_b 
\left(s_b(t + \tau) q_b^{ij} \right) 
\right) 
\right] , 
\label{QQ}
\end{align}
%\begin{align}
%\frac{1}{(4\pi)^2 T_{obs}} \oint d\Omega_a \oint d\Omega_b 
%\int_0^{T_{obs}} dt 
%\left(s_a(t) q_a^{ij} \right)
%\left(s_b(t + \tau) q_b^{ij} \right) , 
%\label{QQ}
%\end{align}
where 
the correlation among pulsar timing data is taken over the full sky 
as well as the whole observation period $T_{obs}$, 
and $\tau$ denotes a time lag in the autocorrelation. 
Here, the subscripts $i$ and $j$ are doubly written but not contracted with each other. 
As a result, 
$q_a^{ij} q_b^{ij}$ is not a tensor but ill-defined 
in the simultaneous limit as $\tau \to 0$. 
Hence, the present paper does not employ the form of Eq. (\ref{QQ}). 

Therefore, we define another form as 
\begin{align}
Q^{ij} 
\equiv 
\frac{1}{T_{obs}} 
 \int_0^{T_{obs}} dt 
\left[
\left(
\frac{1}{4\pi} 
\oint d\Omega_a 
\left(s_a(t) q_a^{ik} \right)
\right)
\left(
\frac{1}{4\pi} 
\oint d\Omega_b 
\left(s_b(t + \tau) q_b^{kj} \right) 
\right) 
\right] , 
\label{Q}
\end{align}
%\begin{align}
%Q^{ij} 
%\equiv 
%\frac{1}{(4\pi)^2 T_{obs}} \oint d\Omega_a \oint d\Omega_b 
%\int_0^{T_{obs}} dt 
%\left(s_a(t) q_a^{ik} \right)
%\left(s_b(t + \tau) q_b^{kj} \right) , 
%\label{Q}
%\end{align}
where the summation is only for $k$ as $\sum_{k=1}^3$. 
Note that $q_a^{ik} q_b^{kj}$ is a tensor 
in the simultaneous limit as $\tau \to 0$. 

We suppose that noises are random in space and time to follow 
\begin{align}
\oint d\Omega_a 
n_a(t) 
&= 0 , 
\notag\\
\oint d\Omega_a 
n_a(t) \hat\Omega_a
&= 0 , 
\notag\\
\oint d\Omega_a 
n_a(t) q_a^{ij} 
&= 0 , 
\notag\\
\int_0^{T_{obs}} dt \:
n_a(t) 
& = 0 .
\label{noise}
\end{align}
The second and third equations in Eq. (\ref{noise}) mean that 
there are no dipole and quadrupole moments of noises. 
By using Eqs. (\ref{sa}), (\ref{sb}) and (\ref{noise}), 
Eq. (\ref{Q}) becomes 
\begin{align}
Q^{ij} 
= 
\frac{1}{T_{obs}} 
 \int_0^{T_{obs}} dt 
 \left[
\left(
\frac{1}{4\pi} 
\oint d\Omega_a 
\left(z_a(t) q_a^{ik} \right)
\right)
\left(
\frac{1}{4\pi} 
\oint d\Omega_b 
\left(z_b(t + \tau) q_b^{kj} \right) 
\right) 
\right] , 
\label{Q2}
\end{align}

The directional average of $F^{A}_{a} q_a^{ij}$ for $A = +, \times$ 
becomes
\begin{align}
\frac{1}{4\pi} \oint d\Omega_a F_{a}^{A} q_a^{ij} 
= C_A e_{A}^{i j}(\hat\Omega_{GW}) ,
\label{Integral-F-q}
\end{align}
where the sum over $A$  is not taken and 
$C_A = 1/6$. 
See  Appendix A for a derivation of Eq. (\ref{Integral-F-q}). 

By using Eqs. (\ref{za}), (\ref{zb}) and (\ref{Integral-F-q}), 
Eq. (\ref{Q2}) is calculated as 
\begin{align}
Q^{ij} 
&= 
\frac{1}{36 T_{obs}} \int_0^{T_{obs}} dt 
\Bigl(
(h^{+})^2 \cos(\omega t) \cos(\omega (t + \tau)) 
e_{+}^{ik}(\hat\Omega_{GW}) e_{+}^{kj}(\hat\Omega_{GW})
\notag\\
&~~~~~~~~~~~~~~~~~~~~~~~~~~
+ 
(h^{\times})^2 \sin(\omega t) \sin(\omega (t + \tau)) 
e_{\times}^{ik}(\hat\Omega_{GW}) e_{\times}^{kj}(\hat\Omega_{GW})
%\right.
\notag\\
&~~~~~~~~~~~~~~~~~~~~~~~~~~
%\left.
+ h^{+} h^{\times} 
\Bigl[\cos(\omega t) \sin(\omega (t + \tau)) 
e_{+}^{ik}(\hat\Omega_{GW}) e_{\times}^{kj}(\hat\Omega_{GW})
\notag\\
&~~~~~~~~~~~~~~~~~~~~~~~~~~~~~~~~~~~~~~
+ \sin(\omega t) \cos(\omega (t + \tau))
e_{\times}^{ik}(\hat\Omega_{GW}) e_{+}^{kj}(\hat\Omega_{GW})\Bigr]
\Bigr) . 
\label{Q3}
\end{align}

We use 
\begin{align}
e_{+}^{ik} (\hat\Omega_{GW}) e_{+}^{kj} (\hat\Omega_{GW}) 
&= P^{ij}, 
\notag\\
e_{\times}^{ik}(\hat\Omega_{GW}) e_{\times}^{kj}(\hat\Omega_{GW}) 
&= P^{ij}, 
\notag\\
e_{+}^{ik}(\hat\Omega_{GW}) e_{\times}^{kj}(\hat\Omega_{GW}) 
&= - \epsilon^{ijk} \hat\Omega_{GW}^k, 
\label{ee}
\end{align}
where  
$\epsilon^{ijk}$ is the Levi-Civita symbol, 
and  
the projection tensor with respect to $\hat\Omega_{GW}$ 
is defined as 
\cite{CreightonBook, MaggioreBook}
\begin{align}
P^{ij} \equiv \delta^{ij} - \hat\Omega_{GW}^i \hat\Omega_{GW}^j .
\label{P}
\end{align}  
See  Appendix B for the derivation of Eq. (\ref{ee}). 

Substituting Eq. (\ref{ee}) into Eq. (\ref{Q3}) leads to 
\begin{align}
Q^{ij} 
&= 
\frac{1}{36 T_{obs}} \int_0^{T_{obs}} dt 
\Bigl(
%\left(
(h^{+})^2 \cos(\omega t) \cos(\omega (t + \tau)) 
P^{ij} 
+ 
(h^{\times})^2 \sin(\omega t) \sin(\omega (t + \tau)) 
P^{ij} 
%\right.
\notag\\
&~~~~~~~~~~~~~~~~~~~~~~~~~
%\left.
- h^{+} h^{\times} 
\Bigl[\cos(\omega t) \sin(\omega (t + \tau)) 
- 
\sin(\omega t) \cos(\omega (t + \tau))\Bigr]
\epsilon^{ijk} \hat\Omega_{GW}^k
\Bigr) .
%\right) . 
\label{Q4}
\end{align}
Furthermore, we take the limit as $T_{obs} \to \infty$, 
which leads to
$\cos(\omega t) \cos(\omega (t + \tau)) \to \cos(\omega \tau)/2$, 
$\sin(\omega t) \sin(\omega (t + \tau)) \to \cos(\omega \tau)/2$, 
$\cos(\omega t) \sin(\omega (t + \tau)) \to \sin(\omega \tau)/2$, 
and
$\sin(\omega t) \cos(\omega (t + \tau)) \to \sin(\omega \tau)/2$.  
From Eq. (\ref{Q4}), 
we thus arrive at 
\begin{align}
Q^{ij} 
&\to 
\frac{1}{72}
\left( 
\Bigl[(h^{+})^2 + (h^{\times})^2\Bigr]
\cos(\omega \tau) 
P^{ij} 
- 
h^{+} h^{\times} 
\sin(\omega \tau) 
\epsilon^{ijk} \hat\Omega_{GW}^k
\right) . 
\label{Q5}
\end{align}

In addition, we suppose that 
the time lag is much shorter than the period of GWs 
(denoted as $T_{GW}$), 
namely $2\pi (\tau/T_{GW}) = \omega \tau \ll 1$. 
As an example, we can imagine 
$\tau =2$ weeks, and $T_{GW} = 12$ months. 
So, we take the limit as $\omega \tau \to 0$, 
which leads to $\cos(\omega \tau) \to 1$ and $\sin(\omega \tau)  \to 0$. 
Eq. (\ref{Q5}) thus becomes 
\begin{align}
Q^{ij} 
&\to 
K 
P^{ij} , 
\label{Q6}
\end{align}
where 
\begin{align}
K \equiv \frac{1}{72} (h^{tot})^2 , 
\label{K}
\end{align}
and 
$h^{tot} \equiv [(h^{+})^2 + (h^{\times})^2]^{1/2}$ is the GW intensity, 

Some noise might contribute via the trace of $Q^{ij}$. 
In order to avoid such an error contamination from noises, 
it is convenient to consider the traceless part of $Q^{ij}$, 
which can be defined as 
\begin{align}
R^{ij} \equiv 
Q^{ij} - \frac13 Q \delta^{ij} , 
\label{R}
\end{align}
where $Q$ denotes the trace of $Q^{ij}$. 
From Eqs. (\ref{Q6}) and (\ref{R}), 
\begin{align}
R^{ij} 
&\to 
K \left(P^{ij} - \frac13 \delta^{ij} P^{kk} \right) 
\notag\\
&= 
- K q_{GW}^{ij} ,  
\label{R2}
\end{align}
where 
\begin{align}
q_{GW}^{ij} 
\equiv 
\hat\Omega_{GW}^i \hat\Omega_{GW}^j - \frac13 \delta^{ij} . 
\end{align}
Note that $R^{ij}$ is a matrix with rank 3, 
while $P^{ij}$ is a rank-2 matrix.

By calculating the traceless part of Eq. (\ref{Q}), 
$R^{ij}$ is expressed in terms of 
the observed signals $s_a(t)$ and $s_b(t + \tau)$ as 
\begin{align}
R^{ij} 
= 
\frac{1}{(4\pi)^2 T_{obs}} 
\int_0^{T_{obs}} dt 
\oint d\Omega_a \oint d\Omega_b 
\left[ 
s_a(t) s_b(t + \tau)
\left( q_a^{ik} q_b^{kj} - \frac13 \Bigl[(\cos\gamma_{ab})^2 - \frac13 \Bigr] \delta^{ij} \right) 
\right] , 
\label{R3}
\end{align}
where 
$\gamma_{ab}$ is a separation angle between the $a$-th and $b$-th pulsars, 
namely $\cos\gamma_{ab} \equiv \hat\Omega_a \cdot \hat\Omega_b$, 
and we use $q_a^{ik} q_b^{ki} = (\cos\gamma_{ab})^2 - 1/3$.

%%%%
\section{From quadrupole-moment correlations to DOA estimations}
In this section, 
we shall discuss how to estimate $\hat\Omega_{GW}$ 
from the matrix $R^{ij}$.
From Eq. (\ref{R2}), we obtain $\det(R) = - 2 K^3 /27$, 
where 
the determinant of $R^{ij}$ is denoted  as $\det(R)$, 
and 
$\det(q_{GW}^{ij}) = 2/27$ is used. 
This equation is solved for $K$ as 
\begin{align}
K = \left( - \frac{27 \det(R)}{2} \right)^{1/3} . 
\label{K2}
\end{align}

For $i  = x$ and $j = x$, Eq. (\ref{R2}) becomes 
$R^{xx} = - K [(\hat\Omega_{GW}^x)^2 -1/3]$, 
which is solved for $\hat\Omega_{GW}^x$ as 
\begin{align}
\hat\Omega_{GW}^x 
= \pm \sqrt{\frac13 - \frac{R^{xx}}{K}}  .
\label{Omega-x}
\end{align}
There exist both signs of $\pm$, because 
$P^{ij}$ does not distinguish $\hat\Omega_{GW}$ from $- \hat\Omega_{GW}$. 

For $i  = x$ and $j = y$, Eq. (\ref{R2}) reads 
$R^{xy} = - K \hat\Omega_{GW}^x \hat\Omega_{GW}^y$, 
which is rearranged as 
\begin{align}
\hat\Omega_{GW}^y 
= - \frac{R^{xy}}{K \hat\Omega_{GW}^x} , 
\label{Omega-y}
\end{align}
where $\hat\Omega_{GW}^x \neq 0$ is assumed. 
Similarly, Eq. (\ref{R2}) for $i  = z$ and $j = x$ leads to 
\begin{align}
\hat\Omega_{GW}^z 
= - \frac{R^{zx}}{K \hat\Omega_{GW}^x} . 
\label{Omega-z}
\end{align}

From PTA signals, we can estimate $R^{ij}$. 
The components of $R^{ij}$ are substituted into Eqs. (\ref{Omega-x}) -  (\ref{Omega-z}) 
to obtain the GW source direction $\hat\Omega_{GW}$. 

For a unit vector $\hat\Omega_{GW}$,  
at least one of its components, 
$\hat\Omega_{GW}^x$, $\hat\Omega_{GW}^y$ and $\hat\Omega_{GW}^z$ 
never vanishes. 
In a case of $\hat\Omega_{GW}^x = 0$, 
we should use counterparts of Eq. (\ref{Omega-x})-(\ref{Omega-z}). 
If $\hat\Omega_{GW}^y \neq 0$, 
\begin{align}
\hat\Omega_{GW}^y 
&= \pm \sqrt{\frac13 - \frac{R^{yy}}{K}}  .
\label{Omega-y-y}
\\
\hat\Omega_{GW}^z
&= - \frac{R^{yz}}{K \hat\Omega_{GW}^y} , 
\label{Omega-y-z}
\\
\hat\Omega_{GW}^x 
&= - \frac{R^{xy}}{K \hat\Omega_{GW}^y} . 
\label{Omega-y-x}
\end{align}
If $\hat\Omega_{GW}^z \neq 0$, 
\begin{align}
\hat\Omega_{GW}^z 
&= \pm \sqrt{\frac13 - \frac{R^{zz}}{K}}  .
\label{Omega-z-z}
\\
\hat\Omega_{GW}^x
&= - \frac{R^{zx}}{K \hat\Omega_{GW}^z} , 
\label{Omega-z-x}
\\
\hat\Omega_{GW}^y 
&= - \frac{R^{yz}}{K \hat\Omega_{GW}^z} . 
\label{Omega-z-y}
\end{align}
Therefore, the present method can be used in principle for any DOA.

%%%%
\section{Discreteness, noises and estimation errors}
\subsection{From a continuous limit to a discrete distribution of pulsars}
In this section, 
we take account that 
the number of the observed pulsars as well 
as that of the observations are finite. 
We suppose that a finite number ($N_p$) of pulsars are observed 
for a whole observation time $T_{obs}$ with a cadence $P$. 
Each of the pulsars is observed $N_{obs} \equiv [T_{obs}/P] + 1$ times, 
where $[ \quad ]$ denotes a floor function. 

Let us begin with a uniform distribution of pulsars as a particular case. 
Then, 
Eq. (\ref{R3}) with Eq. (\ref{Q2}) can be rewritten in terms of discrete sums as 
\begin{align}
R^{ij} 
=
\frac{1}{{\cal K}_{max} (N_p)^2}
\sum_{K =1}^{{\cal K}_{max}}
\sum_{a =1}^{N_p}
\sum_{b =1}^{N_p}
\left[ 
s_a(t_K) s_b(t_K + \tau)
\left( q_a^{ik} q_b^{kj} 
- \frac13 \Bigl[(\cos\gamma_{ab})^2 - \frac13 \Bigr] \delta^{ij} \right) 
\right] ,
\label{R-sum}
\end{align}
where 
$K$ is a positive integer denoting the $K$-th observation, 
$t_K \equiv P (K - 1)$ denotes the $K$-th observation epoch, 
and 
${\cal K}_{max} \equiv [(T_{obs} - \tau + P)/P]$ 
means the number of the neighboring pair of 
$t = t_K$ and $t = t_K + \tau$ from $t=0$ to $t = T_{obs}$. 
In the rest of this paper, 
we adopt $\tau = P$. 
This leads to ${\cal K}_{max} = N_{obs} -1$.  

In a realistic data analysis, PTA observations are not regular, 
for which there are several sources of timing delays such as clock errors and Solar-system ephemeris errors. 
See \cite{TaylorBook} for a review on timing uncertainties.

PTA data is messy owing to varieties of noises and PTA data points for each pulsar are not regular in time. 
See e.g. \cite{Agazie2023b, Chalumeau} for actual detector characterization and noise budget in PTAs. 
It is expected that random errors cancel out in the linear order level, 
while the second-order term such as the quadruple correlation remains. 
The second-order effect in $R^{ij}$ is $\sim \varepsilon^2/\sqrt{N_{obs}}$, 
where $\varepsilon$ means a fractional error. 
See e.g. Eq. (\ref{deltaR}) for the derivation of the inverse square root of $N_{obs}$.

For example, we suppose the data error or non-regularity in time is about 30 percents 
($\varepsilon \sim 0.3$). 
This implies  two-week cadence for instance may have a fluctuation roughly $\sim 4$ days. 
Then, the effect to $R^{ij}$ is $\sim 0.01$ when $N_{obs} \sim 100$. 
This leads to the DOA estimation error of a few percent,  
below the dominant effects by a random noise $n_a(t)$ 
as will be shown in Section 4.3.

It is noteworthy that Eq. (\ref{R-sum}) is based on the use of the arithmetic mean, 
which is a good approximation in a uniform distribution of 
pulsars. 
For a non-uniform distribution, 
Eq. (\ref{R-sum}) is modified as follows.

\subsection{Weighted averaging for a non-uniform distribution of pulsars}
For a uniform distribution of pulsars, Eq. (\ref{R-sum}) is a discretization  
based on uniform grids. 
We shall consider non-uniform grids 
for the actual distribution of pulsars being non-uniform. 

The surface integral over a unit sphere as $\oint d\Omega_a$ 
can be discretized in the form of a Riemann sum as  
$\sum_a \Delta\Omega_a$, 
where $\Delta\Omega_a$ is the surface area element at $\hat\Omega_a$ 
\cite{Press, Calculus}. 
$\Delta\Omega_a = 4\pi/N_p$ if a distribution of pulsars is uniform. 
On the other hand, 
$\Delta\Omega_a$ is not a constant when the distribution is not uniform. 
One such example is the global surface temperature 
that is the average temperature of Earth's surface at a given time 
\cite{Climate, Climate2}. 

Therefore, we employ an area-weighted average 
that calculates the mean value of a certain variable across a surface. 
Eq. (\ref{R}) is thus discretized as 
\begin{align}
R^{ij} 
=
\frac{1}{{\cal K}_{max}}
\sum_{K =1}^{{\cal K}_{max}}
\sum_{a =1}^{N_p}
\sum_{b =1}^{N_p}
\left[ 
w_a w_b s_a(t_K) s_b(t_K + \tau)
\left( q_a^{ik} q_b^{kj} 
- \frac13 \Bigl[(\cos\gamma_{ab})^2 - \frac13 \Bigr] \delta^{ij} \right) 
\right] ,
\label{R-sum-w}
\end{align}
where $w_a$ and $w_b$ denote normalized weights for $a$-th and $b$-th pulsars, respectively. 
Namely, $w_a = \Delta\Omega_a/(4\pi)$ and $\sum_a w_a = 1$. 
Eq. (\ref{R-sum-w}) recovers Eq. (\ref{R-sum})  when 
a distribution of pulsars is uniform ($w_a = w_b = 1/N_p$).

We use a Voronoi tessellation
(also called as a Voronoi diagram or Voronoi decomposition) 
to numerically compute $w_a$ via $\Delta\Omega_a$. 
The Voronoi tessellation, which has many applications in engineering as well as science, 
is a partition of a plane into regions close to each of a given set of objects 
\cite{VoronoiBook1, VoronoiBook2}. 
For instance, it has been used in cosmology to discuss 
a distribution of galaxies on a large-scale structure of the universe 
e.g. \cite{Coles, Paranjape}, 
which can be dated back to 
\cite{Matsuda1984,Icke1987}. 
Furthermore, the Voronoi tessellation is used extensively for estimating global surface temperature, 
because the spatial distribution of weather stations is very non-uniform on Earth. 
It partitions a world map into cells based on weather station proximity, 
in which every point in a cell is closer to its specific station than to any other.

\subsection{Estimation error from a random noise}
In this subsection, 
we shall discuss a typical size of the estimation error due to random noises,  
where we suppose $n_a(t_K)$ and $n_b(t_{K'})$ are white random noises 
with the zero mean and the common standard deviation $\sigma$. 
Let us write down relations between $n_a(t_K)$ and $n_b(t_{K'})$ for the later convenience.

The zero mean for a weighted average over pulsar directions 
or over the whole observation time is 
\begin{align}
\sum_{a=1}^{N_p} w_a n_a(t_K) 
&\approx 0 ,
\label{n-p}
\\
\frac{1}{{\cal K}_{max}} 
\sum_{K=1}^{{\cal K}_{max}} n_a(t_K) 
&\approx 0 .
\label{n-t}
\end{align}

We assume also that 
there are nearly no correlations between different directions 
or between different observation epochs, 
expressed as  
\begin{align}
\sum_{a \neq b} w_a w_b n_a(t_K) n_b(t_K)
&\approx 0 ,
\label{nn-p}
\\
\sum_{K \neq K'} n_a(t_K) n_a(t_{K'})
&\approx 0 .
\label{nn-t}
\end{align} 
In astronomy, noises in proximity to each other may be correlated. 
Such correlations are reduced in the present method because 
the weights of pulsars in close proximity to each other 
are much smaller than an average one ($1/N_p$). 
Therefore, the present paper uses Eqs. (\ref{nn-p}) and (\ref{nn-t}) as an approximation 
in order-of-magnitude estimations. 

The noises have the common variance $\sigma^2$ as 
\begin{align}
\sum_{a=1}^{N_p} w_a  \left( n_a(t_K) \right)^2 
&\approx \frac{\sigma^2}{N_p} ,
\label{n2-p}
\\
\frac{1}{{\cal K}_{max}} 
\sum_{K=1}^{{\cal K}_{max}} \left( n_a(t_K) \right)^2
&\approx \sigma^2 .
\label{n2-t}
\end{align}

A correlation between noises observed simultaneously 
is calculated as 
\begin{align}
\sum_{a=1}^{N_p} \sum_{b=1}^{N_p}
w_a w_b n_a(t_K) n_b(t_K)
&\approx 
\sum_{a=1}^{N_p} 
\left( w_a n_a(t_K) \right)^2
\notag
\\
&
\lesssim
\sum_{a=1}^{N_p} 
w_a \left( n_a(t_K) \right)^2
\notag
\\
&\approx 
\frac{\sigma^2}{N_p} ,
\label{n2-pp}
\end{align}
where Eqs. (\ref{nn-p}) and (\ref{n2-p}) are used in the first and third lines, respectively, 
and $0 < w_a \lesssim 1$ for the normalized weight is used in the second line.    
Similarly, noises for the same pulsar satisfy 
 \begin{align}
\frac{1}{({\cal K}_{max})^2} 
\sum_{K=1}^{{\cal K}_{max}} \sum_{K' = 1}^{{\cal K}_{max}} 
n_a(t_K) n_a(t_{K'})
&\approx 
\frac{1}{({\cal K}_{max})^2} 
\sum_{K=1}^{{\cal K}_{max}}  
\left( n_a(t_K) \right)^2
\notag
\\
&\approx \frac{\sigma^2}{{\cal K}_{max}} ,
\label{n2-tt}
\end{align}
where Eqs. (\ref{nn-t}) and (\ref{n2-t}) are used in the first and second lines, respectively.

Let us discuss the perturbation of $R^{ij}$ owing to the noises, denoted as 
$\delta_n R^{ij}$, 
which is roughly 
\begin{align}
\delta_n R^{ij} 
\sim 
\left(\frac{1}{{\cal K}_{max}} 
\sum_{K=1}^{{\cal K}_{max}} \sum_{a =1}^{N_p}\sum_{b =1}^{N_p} 
w_a w_b n_a(t_K)  n_b(t_K + P) \right) . 
\label{deltan-R}
\end{align}
Here, we focus on the quadratic terms in noises 
in Eq. (\ref{R-sum}) 
by ignoring the coupling terms of GWs and noises, 
because 
the coupling terms are $\sim h^{tot} \times \sigma$, 
the quadratic noise terms are $\sigma^2$, 
and 
$h^{tot} \ll \sigma$ is expected for the current PTA observations.

Note that the mean of $\delta_n R^{ij}$ vanishes  
owing to Eq. (\ref{n-t}). 
Therefore, the variance of $R^{ij}$ 
is thus calculated as 
\begin{align}
&(\delta_n R^{ij})^2 
\notag\\
\sim& 
\left(\frac{1}{{\cal K}_{max}} 
\sum_{K=1}^{{\cal K}_{max}} \sum_{a =1}^{N_p}\sum_{b =1}^{N_p} 
w_a w_b n_a(t_K)  n_b(t_K + P) \right)
\notag\\
&\times
\left(\frac{1}{{\cal K}_{max}} 
\sum_{K' = 1}^{{\cal K}_{max}} \sum_{c =1}^{N_p} \sum_{d =1}^{N_p} 
w_c w_d n_c(t_{K'})  n_d(t_{K'} + P) \right)
\notag\\
\sim& 
\frac{1}{({\cal K}_{max})^2}
\sum_{K=1}^{{\cal K}_{max}} 
\left( 
\sum_{a =1}^{N_p} \sum_{c =1}^{N_p}
w_a w_c n_a(t_K)  n_c(t_K)  
\right)
\notag\\
&~~~~~~~~~~~~~~~~~~~~~~~
\times
\left( 
\sum_{b =1}^{N_p} \sum_{d =1}^{N_p}
w_b w_d n_b(t_K + P)  n_d(t_K + P)  
\right)
\notag\\
\lesssim &
\frac{1}{({\cal K}_{max})^2}
\sum_{K=1}^{{\cal K}_{max}}
\left(  \frac{\sigma^2}{N_p}  \right)^2 
\notag\\
\sim &
\frac{\sigma^4}{{\cal K}_{max} (N_p)^2} , 
\label{deltaR^2}
\end{align}
where 
Eqs. (\ref{n-t}) and (\ref{nn-t}) are used in the fourth and fifth lines, 
respectively, and 
Eq. (\ref{n2-p}) is used in the sixth line. 
Note that the summation is not taken for $i, j$

By using Eq. (\ref{deltaR^2}), 
the size of $\delta_n R^{ij}$ is obtained as 
\begin{align}
|\delta_n R^{ij}| 
&= 
\sqrt{\left(\delta_n R^{ij}\right)^2} 
\notag\\
&\sim 
\frac{\sigma^2}{N_p  \sqrt{N_{obs}}} , 
\label{deltaR}
\end{align}
where we use ${\cal K}_{max} \approx N_{obs}$ 
for $N_{obs} \gg 1$.

The ratio of $\delta_n R^{ij}$  to $R^{ij}$ 
gives us a typical size of the estimation error for a component of the GW directional vector. 
It is 
\begin{align}
\left| \delta_n \hat{\Omega}^i_{GW} \right| 
&\sim
\left| \frac{\delta_n R^{ij}}{R_{GW}^{ij}} \right| 
\notag\\
&\sim  
\frac{1}{N_p \sqrt{N_{obs}}} 
\times \left(\frac{\sigma} {h^{tot}}\right)^2  ,
\label{deltan}
\end{align}
where $R_{GW}^{ij}$ denotes the purely GW part of $R^{ij}$, 
and 
we use $R_{GW}^{ij} \sim (h^{tot})^2$ 
from Eqs. (\ref{K}) and (\ref{R2}). 
This leads to the typical size of an estimation error in the GW direction as 
\begin{align}
\left| \delta_n \hat{\Omega}_{GW} \right| 
&= 
\sqrt{\left| \delta_n \hat{\Omega}^x_{GW} \right|^2 
+ \left| \delta_n \hat{\Omega}^y_{GW} \right|^2 
+ \left| \delta_n \hat{\Omega}^z_{GW} \right|^2}
\notag\\
&\sim
\sqrt{3} \times 
\sqrt{\left| \delta_n \hat{\Omega}^i_{GW} \right|^2}
\notag\\
&\sim  
\frac{\sqrt{3}}{N_p \sqrt{N_{obs}}} 
\times \left(\frac{\sigma} {h^{tot}}\right)^2  ,
\label{deltan2}
\end{align}

This is evaluated as 
\begin{align}
\left| \delta_n \hat{\Omega}_{GW} \right| 
&\sim  
0.1 
\Bigg( \frac{128}{N_p} \Bigg)
\Bigg( \frac{24}{N_{obs}} \Bigg)^{1/2}  
\Bigg( \frac{\sigma}{10} \Bigg)^2
\Bigg( \frac{\sqrt{2}}{h^{tot}} \Bigg)^2  .
\label{deltan3}
\end{align}
Roughly speaking, the estimation error is 
$\sim 0.1$ radian (corresponding to $\sim 6$ degrees) 
for e.g. $N_p \sim 100$, $N_{obs} \sim 24$, 
$\sigma/h^+ \sim \sigma/h^{\times} \sim 10$.

The above order-of-magnitude estimation is based on the white noise assumption. 
For improving the error estimation,  noise correlations are needed to carefully consider 
\cite{Taylor}. 
Achromatic low-frequency noises arise from e.g. intrinsic instabilities of pulsars.  
Chromatic low-frequency processes are a consequence of propagation of radio pulses 
through interstellar space filled with ionized matter, 
where radio signals at lower frequencies experience greater delays.  
See \cite{TaylorBook} for a review on low-frequency noises and their modelings.

\begin{figure}
\includegraphics[width=7.5cm]{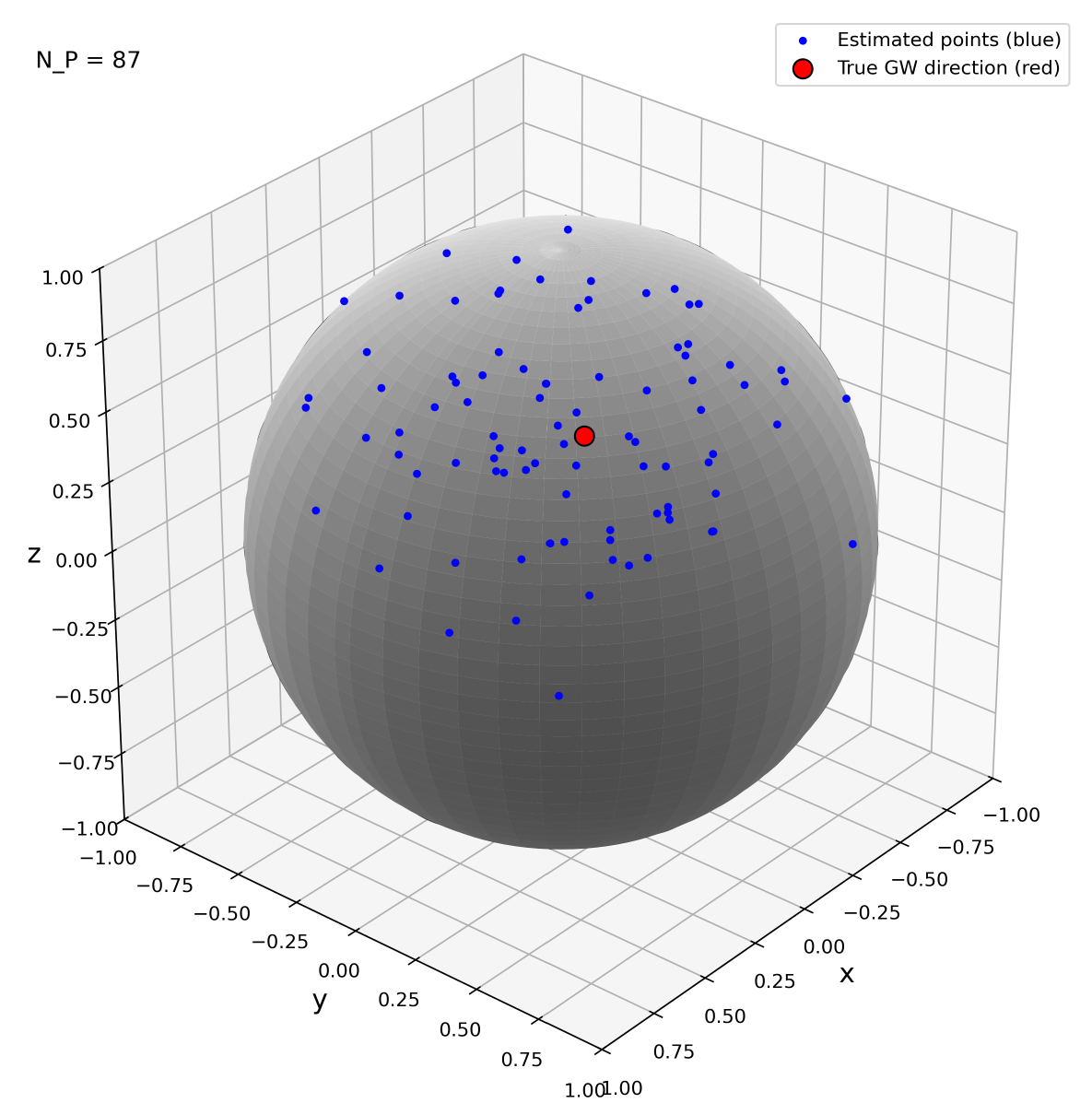}
\includegraphics[width=7.5cm]{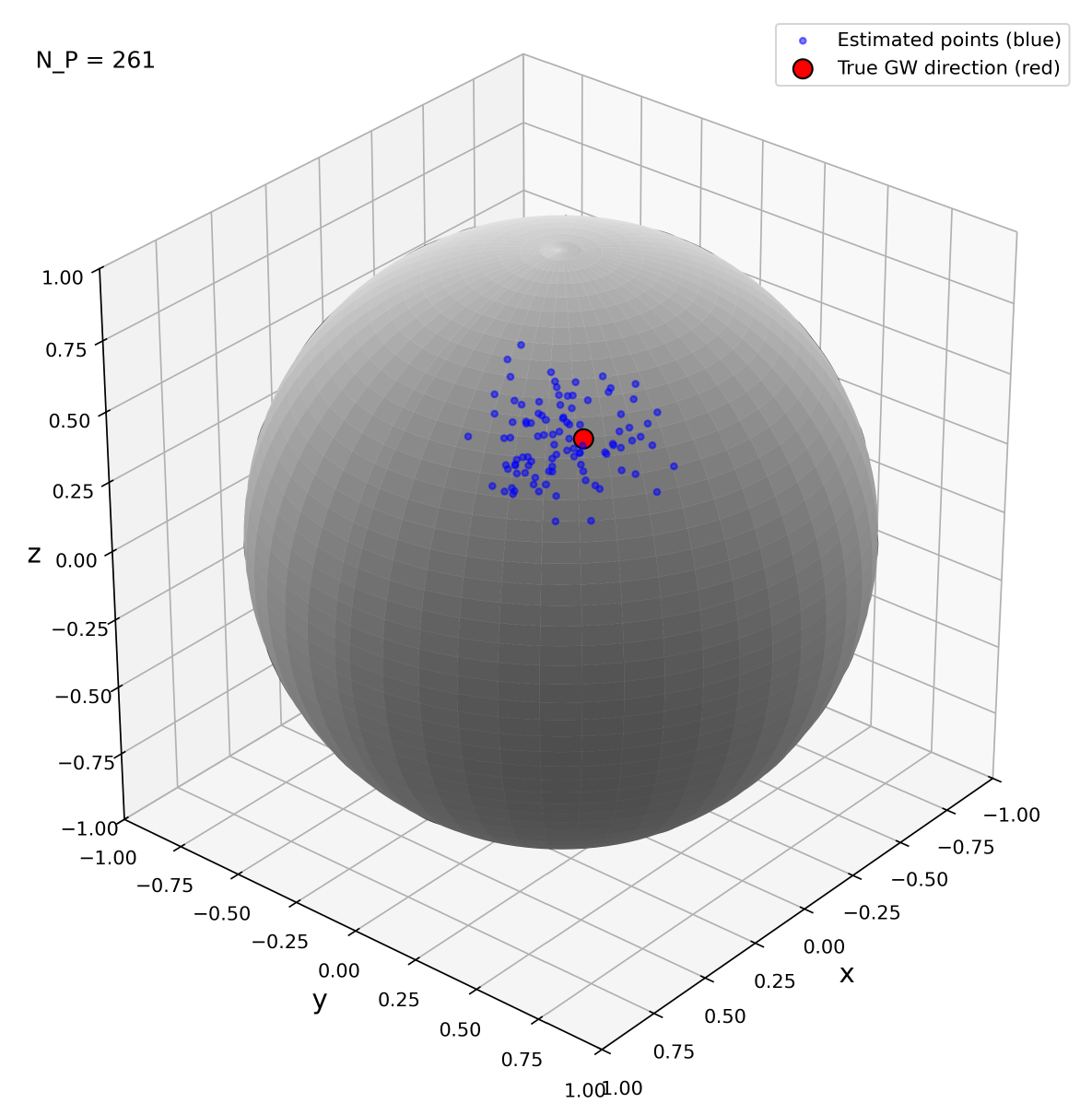}
\caption{
Scatter plots of DOA estimations for a single GW.  
$h^+ = h^{\times} = 1$, and $\sigma = 10$ are chosen 
for $N_{obs} = 260$, 
and 100 runs are numerically performed. 
Here, $\hat{\Omega}_{GW} = (x, y, z)$ is chosen  
as $\theta_{GW} = \phi_{GW} = 45$ deg. 
Left panel: $N_p = 87$, right panel: $N_p = 261$. 
}
\label{fig-scatter}
\end{figure}

Figure \ref{fig-scatter} shows DOA estimation errors depend on 
the number of observed pulsars. 
The left panel is for $N_p = 87$, 
where the pulsars are observed actually by 
NanoGrav, the European Pulsar Timing Array, and Parkes Pulsar Timing Array. 

We assume $h^+ = h^{\times} =1$ for GW amplitudes, 
the Gaussian noise for $n_a(t)$ with $\sigma = 10$, 
and  $N_{obs} = 260$. 
Each realization corresponds to a (blue in color) dot in the figure, 
where one hundred realizations are made. 
The dots in this figure denote the plus branch of Eq. (\ref{Omega-x}), 
whereas estimated points for  the minus branch are located on the other side of the unit sphere 
and they are thus unseen in this figure. 

The right panel is for $N_p = 261$ as a hypothetical case based on the expectation of SKA, 
where $N_p = 261$ is produced by adding two hypothetical pulsars around each of the eighty seven known pulsars. 
The parameters $h^+, h^{\times}, \sigma, N_{obs}$ in the right panel 
are the same as those in the left one. 
We assume that the sky position of the added pulsars 
follows a Gaussian distribution with the standard deviation of $0.1$ radian. 
Why do we adopt $0.1$ radian? 
A spatial distribution of pulsars newly found by SKA 
is likely to associate with the stellar distribution in our galaxy. 
$0.1$ radian corresponds to the apparent size of stellar distributions such as 
a globular cluster ($\sim 100$ pc) at a typical distance ($\sim 1$ kpc) of PTA pulsars. 
Note that $\sigma$ in the SKA era may be smaller than that in the current era. 
In this sense, the scatter in the right panel could be reduced owing to the improved $\sigma$ by SKA.

In the left panel of Figure \ref{fig-scatter}, 
several dots are located at points completely different from the true DOA 
denoted by a red disk. 
This suggests that $N_p = 87$ 
(nearly the current status of PTAs 
\cite{Agazie2023, Antoniadis2023, Reardon2023, Xu2023}) 
is not enough to estimate DOAs 
by using the quadrupole-correlation method.  

The right panel of Figure \ref{fig-scatter} for $N_p = 261$, 
on the other hand, 
suggests that the quadrupole-correlation method could potentially 
allow a DOA estimation in the SKA era.

As mentioned in Introduction, 
there has been a work on localizing binaries on the sky by using PTAs, 
especially through the pulsar term 
that depends on pulsar distance e.g. \cite{Lee, Rosado, Petrov, Taylor2026, Kopeikin}. 
Those papers may suggest that 
inclusion of the pulsar term were necessary for the source localization by PTAs, 
Yet, the current result means that the pulsar term is not always indispensable 
for an estimation of the DOA, 
though the pulsar term is needed to improve the accuracy. 
The quadrupole correlations bring the information about the DOA 
even if the pulsar term is ignored. 
Therefore, the current method cannot be derived from the existing methods. 
The current method is thus independent from 
the existing approaches including the pulsar term. 
In spite of the crucial difference, 
the expected accuracy of the sky area in the both approaches  
is similar to $\sim 10^2$ square degrees \cite{Lee, Rosado, Petrov}.

What limits the current method? 
PTAs are sensitive to GWs across $10^{-9} - 10^{-7}$ Hz frequency band. 
In the current paper, a non-evolutionary binary is considered. 
Chirp effects may change the story. 
PTAs are sensitive to SMBHBs with a mass range from $10^8$ to $10^{10} M_{\odot}$ 
around $10^5$ years before the merger. 
Especially for the merger time less than $\sim 10^3$ years, 
the frequency chirp will be visible in e.g. five-year data \cite{Lee}. 
For such an evolving binary, the current method does not hold as it is.

\subsection{Single dominant source over an isotropic stochastic background}
Finally, let us mention a possible way to make the scenario more realistic. 
We suppose that a signal from a single SMBHB 
coexists with a stochastic GW background (GWB)  
in a certain frequency bin of PTA observations, 
where the signal can be expressed as 
$s_a(t) = z_a(t) + z_{a BG}(t) + n_a(t)$ 
for $ z_{a BG}$ denoting the redshift due to the stochastic background.

We assume the isotropic stochastic background 
possibly generated by inflation, 
for which there can exist quadrupole (and higher) moments of the redshifts 
due to the GWB, 
namely $\oint d\Omega_a z_{a BG}(t) q_{a}^{ij} \neq 0$. 
When the signal from a SMBHB 
is significantly stronger than that from the isotropic stochastic GWB, 
we can expect 
$|\oint  d\Omega_a z_a q_a^{ik}| \gg |\oint  d\Omega_a z_{a BG} q_a^{ik}|$. 
From Eq. (\ref{Q}), therefore, we can arrive at Eq. (\ref{Q6}) as an approximation. 
Namely, the present formulation and method can be used for a strong signal of a SMBHB 
over the stochastic GWB.

%%%%%
\section{Summary}
We discussed correlations between the quadrupole moments of 
pulsar timing data due to GWs, 
where we supposed that an isolated source such as a SMBHB 
dominates over the isotropic stochastic background 
in a certain bin of the PTA frequency domain. 
This situation was formulated to demonstrate that a DOA of a GW can be estimated 
from the quadrupole correlations. 
The present formulation is not derived from assumptions of a priori Gaussian noise.  
Furthermore, it is an analytic solution that does not need a huge template bank. 
Estimation errors in the present method were also studied.

The Voronoi tessellation was employed to numerically calculate a weighed mean 
for a non-uniform distribution of pulsars on the sky. 
According to numerical calculations, 
the forthcoming SKA could play a role in DOA estimations by quadrupole correlations 
as an independent consistency check.

Is the present method applicable also in the presence of a sub-dominant GW component 
in the same frequency bin? 
Do higher multipole moments such as octupole moments play any role 
in improving the DOA estimation? 
These issues are left for future.

%%%%%
\appendix
\section{Derivation of Eq. (\ref{Integral-F-q}) with $C_{A}$}
In this appendix, 
we shall derive Eq. (\ref{Integral-F-q}) by several steps as follows. 

Let us begin with the angular average of PTA antenna pattern functions 
in Eq. (\ref{F}), 
which is defined as 
\begin{align}
\frac{1}{4\pi} \oint d\Omega_a F_{a}^{A}  
= 
\frac{1}{8\pi} e_{i j}^A(\hat\Omega_{GW}) 
\oint d\Omega_a 
\frac{\hat\Omega_{a}^i \hat\Omega_a^j} 
{1+\hat\Omega_{GW} \cdot \hat\Omega_a}  .
\label{Integral-F}
\end{align}
The integral in the right-hand side of Eq. (\ref{Integral-F}) 
is symmetric between $i$ and $j$, 
while the integrand depends on $\hat\Omega_{GW}$. 
Therefore, the integral is a linear combination of 
$\delta^{ij}$ and $\hat\Omega_{GW}^i \hat\Omega_{GW}^j$. 
Therefore,  
\begin{align}
\frac{1}{4\pi} \oint d\Omega_a F_{a}^{A}  
= 0 , 
\label{Integral-F2}
\end{align}
where we use the transverse and traceless property as 
$e_{i j}^A(\hat\Omega_{GW}) \hat\Omega_{GW}^i \hat\Omega_{GW}^j = 0$ 
and 
$e_{i j}^A(\hat\Omega_{GW}) \delta^{ij} = 0$. 

Secondly, we consider the angular average of a dipole part of the antenna pattern,  
defined as 
 \begin{align}
\frac{1}{4\pi} \oint d\Omega_a F_{a}^{A} \hat\Omega_a^k
= 
\frac{1}{8\pi} e_{i j}^A(\hat\Omega_{GW}) 
\oint d\Omega_a 
\frac{\hat\Omega_{a}^i \hat\Omega_a^j \hat\Omega_a^k} 
{1+\hat\Omega_{GW} \cdot \hat\Omega_a}  .
\label{Integral-F-dipole}
\end{align}
The integral in the right-hand side of Eq. (\ref{Integral-F-dipole}) 
is totally symmetric among $i, j, k$, 
and it depends on $\hat\Omega_{GW}$. 
Hence, the integral is proportional 
$\delta^{ij} \hat\Omega_{GW}^k + \delta^{jk} \hat\Omega_{GW}^i + \delta^{ki} \hat\Omega_{GW}^j$. 
Therefore, 
\begin{align}
\frac{1}{4\pi} \oint d\Omega_a F_{a}^{A} \hat\Omega_a^k
= 0 ,
\label{Integral-F-dipole2}
\end{align}
where we use 
$e_{i k}^A(\hat\Omega_{GW}) \hat\Omega_{GW}^k = 0$ 
and 
$e_{i j}^A(\hat\Omega_{GW}) \delta^{ij} = 0$. 

Thirdly, we examine the angular average of a quadrupole part of the antenna pattern,  
defined as  
\begin{align}
\frac{1}{4\pi} \oint d\Omega_a F_{a}^{A} \hat\Omega_a^k \hat\Omega_a^l
= 
\frac{1}{8\pi} e_{i j}^A(\hat\Omega_{GW}) 
\oint d\Omega_a 
\frac{\hat\Omega_{a}^i \hat\Omega_a^j \hat\Omega_a^k \hat\Omega_a^l} 
{1+\hat\Omega_{GW} \cdot \hat\Omega_a}  .
\label{Integral-F-quadrupole}
\end{align}

The integral in the right-hand side of Eq. (\ref{Integral-F-quadrupole}) 
is totally symmetric among $i, j, k, l$, 
and it depends on $\hat\Omega_{GW}$. 
Therefore, the integral is written as a linear combination of three types as 
$\delta^{ij} \delta^{kl}$ (and its permutations), 
$\delta^{ij} \hat\Omega_{GW}^k \hat\Omega_{GW}^l$ (and its permutations), 
and 
$\hat\Omega_{GW}^i \hat\Omega_{GW}^j \hat\Omega_{GW}^k \hat\Omega_{GW}^l$.  
Most of them vanish for $e_{i j}^A(\hat\Omega_{GW})$, 
because they are longitude or a trace part. 
Only the nonvanishing terms are 
$(\delta^{ik}\delta^{jl} + \delta^{il}\delta^{jk}) e_{i j}^A(\hat\Omega_{GW}) 
= 2 e_{kl}^A(\hat\Omega_{GW})$. 
Therefore, we find 
\begin{align}
\frac{1}{4\pi} \oint d\Omega_a F_{a}^{A} \hat\Omega_a^k \hat\Omega_a^l
= C_A  e_{k l}^A(\hat\Omega_{GW}) ,
\label{Integral-F-quadrupole2}
\end{align}
where $C_A$ is a certain constant to be calculated below. 

\begin{figure}
\includegraphics[width=12.0cm]{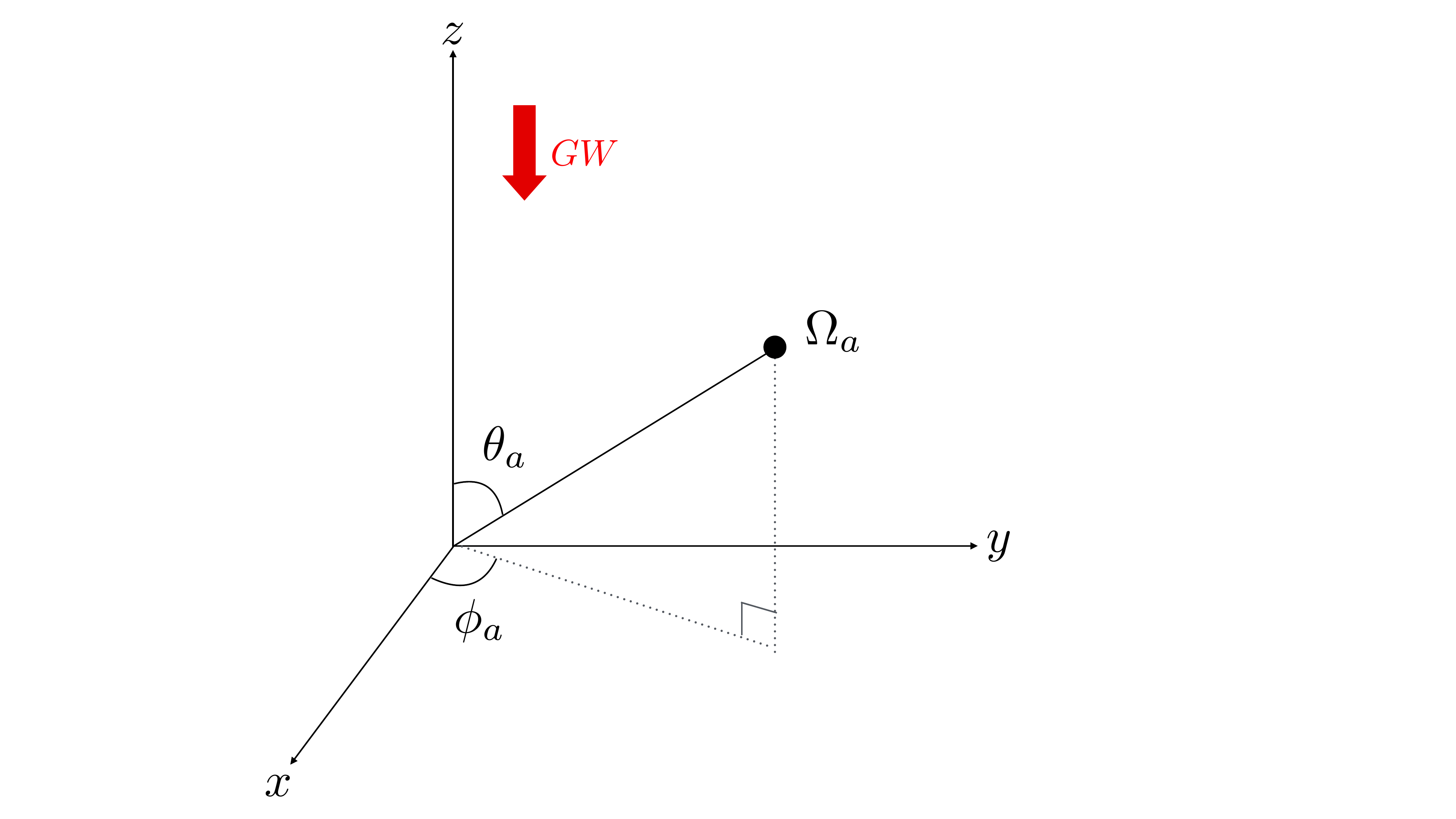}
\caption{
The polar coordinates associated with Eq. (\ref{e}) and $\hat\Omega_{GW} = (0, 0, 1)$. 
}
\label{fig-Omega}
\end{figure}

We choose Cartesian coordinates satisfying 
\begin{align}
e_{i j}^A(\hat\Omega_{GW}) 
= 
\begin{pmatrix}
1 & 0 & 0 \\
0 & -1 & 0 \\
0 & 0 & 0
\end{pmatrix} , 
\label{e} 
\end{align}
where we assume $\hat\Omega_{GW} = (0, 0, 1)$. 
See Figure \ref{fig-Omega} for a schematic figure of $\hat\Omega_{GW}$ 
in the adopted coordinates. 
Note that it is possible to choose $\hat\Omega_{GW} = (0, 0, -1)$, 
because a forward direction or a backward one does not matter 
in this appendix.

For $k =x$ and $l=x$, 
Eq. (\ref{Integral-F-quadrupole2}) becomes
 \begin{align}
\frac{1}{4\pi} \oint d\Omega_a F_{a}^{A} \left(\hat\Omega_a^x\right)^2 
= C_+ ,
\label{Integral-F-quadrupole-xx}
\end{align}
 where Eq. (\ref{e}) is used. 
 $\hat\Omega_a = (\sin\theta_a \cos\phi_a, \sin\theta_a \sin\phi_a, \cos\theta_a)$ in the polar coordinates 
 is substituted into the left-hand side of Eq. (\ref{Integral-F-quadrupole-xx}). 
Straightforward calculations of the left-hand side in Eq. (\ref{Integral-F-quadrupole-xx}) lead to 
\begin{align}
\frac{1}{4\pi} \oint d\Omega_a 
\frac{\sin\theta_a \cos(2\phi_a) (\sin\theta_a \cos\phi_a)^2}{1 + \cos\theta_a}
= \frac16 .
\label{C+}
\end{align}
Therefore, $C_+ = 1/6$. 
Similarly, we obtain $C_{\times} = 1/6$. 
As a result, 
$C_A = 1/6$ for any of $A = +, \times$. 
 
It follows that the trace part of  Eq. (\ref{Integral-F-quadrupole2}) 
leads to Eq. (\ref{Integral-F2}), because $\hat\Omega_a$ is a unit vector. 
Hence, the traceless part of Eq. (\ref{Integral-F-quadrupole2}) 
becomes Eq. (\ref{Integral-F-q}), 
because $q_{ab}^{kl}$ is the traceless part of $\hat\Omega_a^k \hat\Omega_a^l$ 
and $e_{k l}^A(\hat\Omega_{GW})$ is traceless. 

An alternative and straightforward method for deriving Eq. (\ref{Integral-F-q}) 
is to adopt the coordinates associated with Eq. (\ref{e}) when we calculate $C_A$. 
Then, the left-hand side of Eq. (\ref{Integral-F-q}) for $A = +$ 
is directly calculated as 
$1/6$ ($k=x, l=x$), $-1/6$ ($k =y, l=y$),  
and $0$ (otherwise), 
while that for $A = \times$ 
is $1/6$ ($k=x, l=y$ and $k=y, l=x$) and $0$ (otherwise). 
Expressions for these results lead to Eq. (\ref{Integral-F-q}) again.

\section{Relationship between quadratic polarization tensors} 
We adopt the spatial orthonormal vectors $\hat{l}$, $\hat{m}$, $\hat{n}$ 
in the right-handed system, 
where $\hat{l}$ denotes the direction of GW propagation. 
The polarization tensors $e^A_{ij}(\hat\Omega_{GW})$ can be written as 
\cite{CreightonBook, MaggioreBook, Anholm2009, Jenet}
\begin{align}
e^+_{ij}(\hat\Omega_{GW})
&= 
\hat{m}_i \hat{m}_j - \hat{n}_i \hat{n}_j , 
\label{e+}
\\
e^{\times}_{ij}(\hat\Omega_{GW})
&= 
\hat{m}_i \hat{n}_j + \hat{n}_i \hat{m}_j .
\label{ecross}
\end{align}
Note that the GW source direction is opposite to the propagation direction, 
namely
$\hat\Omega_{GW} = - \hat{l}$. 
Therefore,  $\hat{m}$, $\hat{n}$, $\hat\Omega_{GW}$ are 
in the left-handed system.

From Eq. (\ref{e+}), we find 
\begin{align}
e^+_{ik}(\hat\Omega_{GW}) e^+_{kj}(\hat\Omega_{GW}) 
&= 
(\hat{m}_i \hat{m}_k - \hat{n}_i \hat{n}_k)
(\hat{m}_k \hat{m}_j - \hat{n}_k \hat{n}_j) 
\notag\\
&= 
\hat{m}_i \hat{m}_j + \hat{n}_i \hat{n}_j 
\notag\\
&=  
\delta_{ij} - \hat\Omega_{GW}^i \hat\Omega_{GW}^j ,
\label{e+e+}
\end{align}
where 
$\delta_{ij}  = \hat{m}_i \hat{m}_j + \hat{n}_i \hat{n}_j + \hat\Omega_{GW}^i \hat\Omega_{GW}^j$ 
is used in the last line. 
By using Eq. (\ref{P}), we obtain 
\begin{align}
e^+_{ik}(\hat\Omega_{GW}) e^+_{kj}(\hat\Omega_{GW}) 
= P_{ij} . 
\label{e+e+}
\end{align}

From Eq. (\ref{ecross}), we obtain 
\begin{align}
e^{\times}_{ik}(\hat\Omega_{GW}) e^{\times}_{kj}(\hat\Omega_{GW}) 
&= 
(\hat{m}_i \hat{n}_k + \hat{n}_i \hat{m}_k)
(\hat{m}_k \hat{n}_j + \hat{n}_k \hat{m}_j) 
\notag\\
&= 
\hat{m}_i \hat{m}_j + \hat{n}_i \hat{n}_j 
\notag\\
&=  
\delta_{ij} - \hat\Omega_{GW}^i \hat\Omega_{GW}^j ,
\label{e+e+}
\end{align}
where 
$\delta_{ij}  = \hat{m}_i \hat{m}_j + \hat{n}_i \hat{n}_j + \hat\Omega_{GW}^i \hat\Omega_{GW}^j$ 
is used in the last line. 
By using Eq. (\ref{P}), we obtain 
\begin{align}
e^{\times}_{ik}(\hat\Omega_{GW}) e^{\times}_{kj}(\hat\Omega_{GW}) 
= P_{ij} . 
\label{ecrossecross}
\end{align}

From Eqs. (\ref{e+}) and (\ref{ecross}), 
we compute 
\begin{align}
e^+_{ik}(\hat\Omega_{GW}) e^{\times}_{kj}(\hat\Omega_{GW}) 
&= 
(\hat{m}_i \hat{m}_k - \hat{n}_i \hat{n}_k)
(\hat{m}_k \hat{n}_j + \hat{n}_k \hat{m}_j) 
\notag\\
&= 
\hat{m}_i \hat{n}_j - \hat{n}_i \hat{m}_j .
\label{e+ecross}
\end{align}

In the left-handed system, 
$\hat\Omega_{GW} 
= - \hat{\ell} 
= - (\hat{m} \times \hat{n})$, 
which is rewritten as 
$\hat\Omega_{GW}^i = - \epsilon^{ijk} \hat{m}_j \hat{n}_k$. 
This leads to 
$\epsilon^{ijk} \hat\Omega_{GW}^k = \hat{m}_i \hat{n}_j - \hat{n}_i \hat{m}_j$. 
By using this for the last line in Eq. (\ref{e+ecross}), 
we arrive at 
\begin{align}
e^+_{ik}(\hat\Omega_{GW}) e^{\times}_{kj}(\hat\Omega_{GW}) 
= - \epsilon^{ijk} \hat\Omega_{GW}^k . 
\label{e+ecross2}
\end{align}

%\section{From $R^{ij}$ to $\hat\Omega_{GW}$}

\acknowledgments
We are grateful to anonymous referees for the useful comments on the earlier version of the manuscript. 
We thank Keitaro Takahashi for the information on the sky positions of 
sixty-five pulsars observed by NanoGrav, EPTA and PPTA. 
We are grateful to Shun Yamamoto, 
Mariko Nomura, and Yuuiti Sendouda 
for useful discussions. 
We wish to thank JGRG34 workshop participants for stimulating conversations. 
This work was supported 
in part by Japan Society for the Promotion of Science (JSPS) 
Grant-in-Aid for Scientific Research, 
No. 24K07009(H.A.).

\end{document}